\documentclass[12pt,epsf,amssymb,ulem,qsymbols]{article}
\usepackage{tabularx}
\usepackage{array}
\usepackage{graphics}
\usepackage{graphicx}
\usepackage{psfrag}
\usepackage{epsfig}
\usepackage{amsmath}
\usepackage{amssymb}
\usepackage{ulem}
\usepackage{setspace}
\usepackage{rotating}
\usepackage{colortbl}
\usepackage{tabularx}
\usepackage{longtable}
\usepackage{slashed}
\usepackage{lineno}
\makeatletter
\usepackage{textcomp}
\usepackage[usenames,dvipsnames]{xcolor}
\usepackage{relsize}

\usepackage{verbatim}

\setlongtables

\newcommand{\msbar}{{\overline{\rm MS}}}

\newcommand{\bea}{\begin{eqnarray}}
\newcommand{\eea}{\end{eqnarray}}
\newcommand{\beq}{\begin{equation}}
\newcommand{\eeq}{\end{equation}}
\newcommand{\ec}{\end{center}}
\newcommand{\bc}{\begin{center}}

\newcommand{\tev}{{\rm TeV}}
\newcommand{\gev}{{\rm GeV}}
\newcommand{\mev}{{\rm MeV}}

\newcommand{\pdir}{p\kern -5.2pt\raise 0.2ex\hbox {/}}

\newcommand{\vdir}{v\kern -5.75pt\raise 0.15ex\hbox {/}}
\newcommand{\kdir}{k\kern -5.75pt\raise 0.15ex\hbox {/}}
\newcommand{\epsdir}{\epsilon\kern -5.0pt\raise 0.15ex\hbox {/}}
\newcommand{\bvdir}{\bar{v}\kern -5.75pt\raise 0.15ex\hbox {/}}
\newcommand{\Ddir}{D\kern -7.75pt\raise 0.20ex\hbox {/}}
\newcommand{\Adir}{A\kern -7.75pt\raise 0.20ex\hbox {/}}
\newcommand{\ldir}{l\kern -5.0pt\raise 0.2ex\hbox{/}}
\newcommand{\varepsdir}{\varepsilon\kern -5.5pt\raise 0.15ex\hbox{/}}

\newcommand{\nn}{\nonumber}
\makeatother

\begin{document}
\unitlength = 1mm

\thispagestyle{empty} 
\begin{flushright}
\begin{tabular}{l}
{\tt \footnotesize LPT 15-57}\\
\end{tabular}
\end{flushright}
\begin{center}
\vskip 3.4cm\par
{\par\centering \textbf{\LARGE  
\Large \bf Lepton Flavor Violation in Exclusive $b\to s$ Decays}}\\
\vskip 1.05cm\par
{\scalebox{.85}{\par\centering \large  
\sc Damir Be\v{c}irevi\'c$^a$, Olcyr Sumensari$^{a,b}$ and Renata Zukanovich Funchal$^{b}$}
{\par\centering \vskip 0.65 cm\par}
{\sl 
$^a$~Laboratoire de Physique Th\'eorique (B\^at.~210)\\
CNRS and Univ. Paris-Sud, Universit\'e Paris-Saclay, 91405 Orsay cedex, France.}\\
{\par\centering \vskip 0.25 cm\par}
{\sl 
$^b$~Instituto de F\'isica, Universidade de S\~ao Paulo, \\
 C.P. 66.318, 05315-970 S\~ao Paulo, Brazil.}\\

{\vskip 1.65cm\par}}
\end{center}

\vskip 0.55cm
\begin{abstract}
Starting from the general effective hamiltonian relevant to the $b\to s$ transitions, we derive the expressions for the full angular distributions of the 
$B  \to K^{(\ast )} \ell_1 \ell_2$ decay modes, as well as for ${\cal B}(B_s \to \ell_1\ell_2 )$ ($\ell_1\neq \ell_2$). 
We point out the differences in the treatment of the lepton flavor violating modes with respect to the lepton flavor conserving ones. 
Concerning the relevant Wilson coefficients we evaluate them in two different scenarios: (i) The (pseudo-)scalar coefficients are obtained using the (pseudo-)scalar coupling 
extracted from the experimental non-zero value of ${\cal B}(h\to \mu\tau)$, (ii) Revisiting a $Z^\prime$-model in which the flavor changing neutral couplings are allowed.  
We provide the numerical estimates of the branching fractions of the above-mentioned modes in both scenarios. 
\end{abstract}
\newpage
\setcounter{page}{1}
\setcounter{footnote}{0}
\setcounter{equation}{0}
%%%%%%%%%%%%%%%%%%%%%%%%%%%%%%%%%%%%%%%%
\noindent

\renewcommand{\thefootnote}{\arabic{footnote}}
%\linenumbers

\setcounter{footnote}{0}
%%%%%%%%%%%  Section 1
\section{\label{sec-0}Introduction}
With the discovery of Higgs boson at the LHC the Standard Model (SM) has become a complete theory describing all known phenomena at the energies around and below the electroweak scale. 
The quest for physics beyond the SM (BSM) is of major importance in order to solve the hierarchy and flavor problems. In that respect the processes mediated by the flavor changing neutral currents (FCNC) 
are particularly interesting because they provide us with a window to BSM physics through low energy experiments. Among those, the most attention in recent years has been devoted to the exclusive $b\to s$ transitions because of the detection and measurement of ${\cal B}(B_s\to \mu^+\mu^-)$  at LHC~\cite{CMS:2014xfa}, in addition to the detailed angular distributions of the $B\to K^{(\ast )} \ell^+\ell^-$~\cite{Aaij:2015oid} and $B_s\to \phi \ell^+\ell^-$~\cite{Aaij:2015esa} decays which gave us access to a number of observables, including those that 
are only mildly sensitive to hadronic uncertainties, while being highly sensitive to the potential effects of BSM physics~\cite{Becirevic:2011bp}. Currently a couple of discrepancies have been observed~\cite{Altmannshofer:2014rta,Descotes-Genon:2015uva} but their interpretation is still 
a subject of controversies which are mostly related to various sources of hadronic uncertainties, cf. e.g.~\cite{Ciuchini:2015qxb}. 

Although the significant effects of BSM physics were expected to affect the hadronic part of the $b\to s\ell^+\ell^-$ processes, it turned out that the most surprising effect came from 
the ratio 
\bea\label{eq:rk}
R_K= { {\cal B}(B\to K \mu^+ \mu^- )_{q^2\in [1,6]\ \gev^2}\over {\cal B}(B\to K e^+ e^-)_{q^2\in [1,6]\ \gev^2}},
\eea
the measured value of which, $R_K =0.745\left(^{+90}_{-74}\right)(36)$~\cite{Aaij:2014ora}, turned out to be $2.6\sigma$ lower than the one predicted in the SM~\cite{Bobeth:2007dw}. Importantly, in this ratio the hadronic uncertainties cancel 
to a very large extent and the discrepancy is then naturally attributed to the violation of the lepton flavour universality. There have been several attempts to describe this discrepancy in terms of various BSM models~\cite{Alonso:2014csa,Becirevic:2015asa}.  
Most of the models allowing to accommodate the lepton flavor universality violation also allow for the lepton flavour violation (LFV). Although the LFV exclusive decays based on $b\to s\ell_1\ell_2$ ($\ell_{1,2} \in \{e,\mu,\tau\}$) 
have not been studied at the LHC so far,~\footnote{A notable exception has been the search for $B_s\to e\mu$ mode at LHCb~\cite{Aaij:2013cby}.} a recent report by CMS on the observation of a $2\sigma$ excess of $h\to\mu\tau$ decay~\cite{Khachatryan:2015kon} 
boosted the interest in studying  $B_s\to \ell_1\ell_2$, $B\to K^{(\ast )} \ell_1\ell_2$, and $B_s\to \phi \ell_1\ell_2$~\cite{Glashow:2014iga,Lee:2015qra,Crivellin:2015era}.

In this paper we provide the explicit expressions for angular distributions and decay rates of the above exclusive processes, in the case of $\ell_1 \neq \ell_2$. As we shall see some of the operators that do not contribute 
to the lepton flavor conserving processes (LFC) can significantly contribute to the LFV ones. By taking the limit $m_1=m_2$ we retrieve the known expressions for the LFC processes. Our formulas are obviously applicable to any similar process and are written in terms of hadronic matrix elements of the relevant operators and the associated Wilson coefficients. To get the Wilson coefficients in the LFV case we will proceed in two ways: 
(i) We will first assume LFV to be generated through the scalar operator, via coupling to the Higgs boson, and estimate the size of the Wilson coefficients $C_{S,P}^{\mu\tau}$ from the experimental information on ${\cal B}(h\to \tau\mu)$, and then predict the decay rates of the above-mentioned processes; (ii) We use a model with a $Z^\prime$-boson in which the LFV is generated by the vector interaction, estimate the Wilson coefficients $C_{9,10}^{\mu\tau}$ by the known information about the $B_s-\overline B_s$ mixing and the other low energy observables. This latter option has been discussed in ref.~\cite{Crivellin:2015era} which we briefly revisit. 

The remainder of this paper is organized as follows: In Sec.~\ref{sec:2} we set the definitions of the effective Hamiltonian, recall the standard parametrization of hadronic matrix elements and derive the formulas for all three types of the exclusive $b\to s \ell_1\ell_2$ decay modes. In Sec.~\ref{sec:3} we discuss the case of the LFV contributions arising from the scalar operator and derive the upper bounds on the specific decay modes using $C_{S,P}$ extracted from  ${\cal B}(h\to \tau\mu)$. In Sec.~\ref{sec:4} we revisit  the upper bounds on the same processes derived in the framework of the $Z^\prime$-model. We briefly summarize in Sec.~\ref{sec:5}.

\section{\label{sec:2}Exclusive $b\to s  \ell_1\ell_2$ decays}

As a starting point we will extend the usual effective Hamiltonian for the $b\to s$ transitions by including the LFV operators
\begin{equation}
\label{heff}
\mathcal{H}_{\mathrm{eff}}=-\frac{4 G_F}{\sqrt{2}}V_{tb}V_{ts}^* \sum_{i=7,9,10,S,P} \Bigg{(}C_i(\mu)\mathcal{O}_i(\mu)+C_i^\prime(\mu)\mathcal{O}_i^\prime(\mu))\Bigg{)},
\end{equation}
where the relevant operators are defined by
\begin{align}
\mathcal{O}_{9} &=\frac{e^2}{g^2}(\bar{s}\gamma_\mu P_L b)(\bar{\ell_1}\gamma^\mu\ell_{2}),& \mathcal{O}_{10}& = \frac{e^2}{g^2}(\bar{s}\gamma_\mu P_L b)(\bar{\ell_1}\gamma^\mu\gamma^5\ell_{2}),\nn \\
\mathcal{O}_{S} &= \frac{e^2}{(4\pi)^2}(\bar{s}P_R b)(\bar{\ell_1}\ell_{2}),& \mathcal{O}_{P} &= \frac{e^2}{(4\pi)^2}(\bar{s}P_R b)(\bar{\ell_1}\gamma_5\ell_{2}), 
\end{align}
and the operators with flipped chirality $\mathcal{O}^\prime_{9,10,S,P}$ are obtained from  $\mathcal{O}_{9,10,S,P}$ by replacing $P_L \leftrightarrow P_R$, where $P_{L/R}=\frac{1}{2}(1\mp \gamma_5)$.   
In the SM the operators ${\cal O}_{9,10}$ play the major role, together with the electromagnetic penguin operator ${\cal O}_7=(e/g^2) m_b(\bar{s}\sigma_{\mu\nu}P_R b)F^{\mu\nu} $. 
The corresponding Wilson coefficients are obtained through a perturbative matching between the full and effective theories at the weak interaction scale $\mu\simeq m_W$ and then run down to the scale 
at which the process takes place, namely $\mu =m_b$. After appropriately absorbing the effects of ${\cal O}_{1-6}$ in the effective Wilson coefficients, one finally has $C_7=-0.304$, $C_9=4.211$, $C_{10}=-4.103$~\cite{Bobeth:1999mk}. 
Other Wilson coefficients in the SM are zero, $C_{7,9,10}^\prime =0$, $C_{S,P}^{(\prime )}=0$. Of course, if $m_1 \neq m_2$ all the Wilson coefficients are zero in the SM and in order to 
generate their non-zero values one needs to work in a specific framework of BSM physics. Before embarking on that part of the problem, we will now derive the expressions for the decay rates and angular distributions (when possible) starting from the Hamiltonian~(\ref{heff}).

\subsection{Leptonic Decay $B_s\to \ell_1 \ell_2$}

We first focus on the simplest exclusive $b\to s\ell_1\ell_2$ mode, $B_s \to  \ell_1\ell_2$, which is also instructive as far as the operators contributing to the process are concerned. 
Of course, and after the trivial replacements, the same expressions will be valid for $B_d\to  \ell_1\ell_2$. We use the standard decomposition of the hadronic matrix element, 
\begin{equation}
\langle 0 | \bar{b} \gamma_\mu \gamma_5 s | B_s(p) \rangle = i p_\mu f_{B_s},
\end{equation}
where $f_{B_s}$ is the $B_s$-meson decay constant, and obtain
\begin{align}
\label{Bsformula}
&
{\cal B}(B_s\to \ell_1^- \ell_2^+)^{\rm theo} =\dfrac{\tau_{B_s}}{64 \pi^3}\frac{\alpha^2 G_F^2}{m_{B_s}^3}  f_{B_s}^2 |V_{tb}V_{ts}^*|^2\lambda^{1/2}(m_{B_s},m_1,m_2)\nn \\
&\qquad\times \Bigg{\lbrace}[m_{B_s}^2-(m_1+m_2)^2]\cdot\left|(C_9-C_9')(m_1-m_2)+(C_S-C_S')\frac{m_{B_s}^2}{m_b+m_s}\right|^2\nn\\
&\hspace{1cm}+[m_{B_s}^2-(m_1-m_2)^2]\cdot\left|(C_{10}-C_{10}')(m_1+m_2)+(C_P-C_P')\frac{m_{B_s}^2}{m_b+m_s}\right|^2\Bigg{\rbrace},
\end{align}
where  $\lambda(a,b,c)=[a^2-(b-c)^2][a^2-(b+c)^2]$. 
%In what follows, throughout this paper, we will sum over the charge conjugated leptons,
%\bea
%{\cal B}(B_s\to \ell_\alpha \ell_\beta ) = {\cal B}(B_s\to \ell_\alpha^+ \ell_\beta^- ) + {\cal B}(B_s\to \ell_\alpha^- \ell_\beta^+ )  .
%\eea
What immediately becomes evident from eq.~(\ref{Bsformula}) is that in the LFV channel the lepton vector current is not conserved, $i\partial_\mu(\bar{\ell}_1 \gamma^\mu \ell_2)=(m_2-m_1)\bar{\ell}_1 \ell_2\neq 0$, and 
the contribution of $C_9^{(\prime )}$ cannot be neglected. Quite obviously, in the limit $m_1 =m_2$ one finds the usual expression for ${\cal B}(B_s\to \ell^+ \ell^-)$. Finally, when confronting theory with the experimental measurements one needs to account for the effect of oscillations in the $B_s-\overline B_s$ system because the time dependence of the $B_s$-decay rate has been integrated in experiment. Therefore, and to a good approximation, one can identify~\cite{DeBruyn:2012wj}
\bea
{\cal B}(B_s\to \ell_1 \ell_2 )_{\rm exp}\approx  {1\over 1-y_s} {\cal B}(B_s\to \ell_1 \ell_2 )^{\rm th}\,,
\eea
where $y_s=\Delta \Gamma_{B_s}/(2 \Gamma_{B_s}) =0.061(9)$, as measured at LHCb~\cite{Aaij:2014zsa}. Notice that the non-conservation of the vector current induces the term in eq.~(\ref{Bsformula}) proportional to $C_9-C_9^\prime$ which involves the difference of the lepton masses and therefore the decay modes $B_s\to \ell_1^- \ell_2^+$ and $B_s\to \ell_1^+ \ell_2^-$ should be studied separately, unless there is a reason that $(C_9-C_9')_{12} = - (C_9-C_9')_{21}$. 
One should therefore be careful in relating the LFV with the LFC contributions via a multiplicative factor. For that to be plausible, one should make sure the contribution proportional to $C_9-C_9'$ in the LFV case is absent. 

\subsection{$B\to K \ell_1 \ell_2$}

Throughout this paper we will use the kinematics of ref.~\cite{Korner:1989qb}, which for the case of $B\to K \ell_1^- \ell_2^+$ means that the main decay axis $z$ is defined in the rest frame of $B$, so that $K$ and the lepton 
pair travel in the opposite directions. The angle between the negatively charged lepton and the decay axis (opposite to the direction of flight of the kaon) is denoted by $\theta_\ell$ and is defined in the lepton-pair rest frame. 
Concerning the hadronic matrix elements we use the following (standard) parametrizations: 
\begin{align}
\langle \bar K(k)|\bar{s}\gamma_\mu b|\bar B(p)\rangle &= \Big{[}(p+k)_\mu- \frac{m_B^2-m_K^2}{q^2}q_\mu \Big{]}f_+(q^2)+\frac{m_B^2-m_K^2}{q^2} q_\mu f_0(q^2),\\
\langle \bar K(k)|\bar{s}\sigma_{\mu\nu} b|\bar B(p)\rangle &= -i (p_\mu k_\nu-p_\nu k_\mu)\frac{2 f_T(q^2,\mu)}{m_B+m_K},
\end{align}
where $f_{+,0,T}(q^2)$ are the hadronic form factors, functions of $q^2 = (p-k)^2=(p_1+p_2)^2$, with $(m_1 +m_2)^2 \leq q^2 \leq (m_B-m_K)^2$. In what follows the scale $\mu=m_b$ will be assumed.
Using the above definitions we can then write the differential decay rate in the following form, 
\begin{align}
\label{eq:semilepPP}
\dfrac{\mathrm{d}{\cal B}}{\mathrm{d}q^2}(\bar B \to \bar K \ell_1^- \ell_2^+)& = \vert{\cal N}_{K}(q^2)\vert^2\times\Big\lbrace 
\varphi_7(q^2) |C_7+C_{7}'|^2 + \varphi_{9}(q^2) |C_{9}+C_{9}'|^2  +  \varphi_{10}(q^2) |C_{10} +C_{10}'|^2   \nonumber \\
& + \varphi_S(q^2) |C_S+C_{S}'|^2+ \varphi_P(q^2) |C_P+C_{P}'|^2 + \varphi_{79}(q^2) \mathrm{Re}[(C_7+ C_7^\prime) (C_{9}+ C_9^\prime)^*]  \nonumber \\
&+ \varphi_{9S}(q^2) \mathrm{Re}[(C_{9}+ C_9^\prime) (C_S+ C_S^\prime)^*]+ \varphi_{10P}(q^2) \mathrm{Re}[(C_{10} +C_{10} ^\prime)(C_P+C_P^\prime)^*]  \Big\rbrace ,
\end{align}
where $\varphi_{i}(q^2)$ depend on kinematical quantities and on the form factors, or more explicitly:~\footnote{In the notation used to write the formulas for $\varphi_{a(b)}(q^2)$ the upper signs correspond to $\varphi_a(q^2)$ and lower to $\varphi_b(q^2)$.}  
\begin{align}
\label{eq:C910}
\varphi_{7}(q^2) &=  \frac{2 m_b^2|f_T(q^2)|^2}{(m_B+m_K)^2} \lambda(m_B,m_K,\sqrt{q^2})\left[1-\frac{(m_1-m_2)^2}{q^2}-\frac{\lambda(\sqrt{q^2},m_1,m_2)}{3 q^4}\right], \nn \\
\varphi_{9(10)}(q^2)&=\frac{1}{2}|f_0(q^2)|^2(m_1\mp m_2)^2 \frac{(m_B^2-m_K^2)^2}{q^2} \left[1-\frac{(m_1\pm m_2)^2}{q^2}\right] \nn \\
&+\frac{1}{2}|f_+(q^2)|^2 \lambda(m_B,m_K,\sqrt{q^2})\left[1-\frac{(m_1\mp m_2)^2}{q^2}-\frac{\lambda(\sqrt{q^2},m_1,m_2)}{3 q^4}\right] \nn ,\\
\varphi_{79}(q^2)&=\frac{2 m_b f_+(q^2)f_T(q^2)}{m_B+m_K} \lambda(m_B,m_K,\sqrt{q^2})\left[ 1-\frac{(m_1-m_2)^2}{q^2}-\frac{\lambda(\sqrt{q^2},m_1,m_2)}{3 q^4}\right],\nn \\
\varphi_{S (P)}(q^2)&=\frac{q^2 |f_0(q^2)|^2}{2(m_b-m_s)^2}(m_B^2-m_K^2)^2 \left[1-\frac{(m_1\pm m_2)^2}{q^2}\right], \nn \\
\varphi_{10P (9S)}(q^2)&=\frac{|f_0(q^2)|^2}{m_b-m_s}(m_1\pm m_2)(m_B^2-m_K^2)^2\left[1-\frac{(m_1 \mp m_2)^2}{q^2}\right].
\end{align}
Finally, the normalization factor in eq.~(\ref{eq:semilepPP}) reads
\begin{equation}
\vert {\cal N}_{K}(q^2)\vert^2=\tau_{B_d}\dfrac{\alpha^2 G_F^2 |V_{tb}V_{ts}^*|^2}{512 \pi^5 m_B^3}\dfrac{\lambda^{1/2}(\sqrt{q^2},m_1,m_2)}{q^2}\lambda^{1/2}(\sqrt{q^2},m_B,m_K).
\end{equation}
Like in the previous subsection we see that due to the non-conservation of the leptonic vector current, the new pieces emerge in the functions $\varphi_i(q^2)$. 
By taking the limit $m_1 = m_2$ in eq.~(\ref{eq:C910}) we retrieve the known expressions for the LFC case. 
We should also emphasize that the interference term $\varphi_{9S}(q^2)$ changes the sign depending on the charge of the heavier lepton. In other words, if one assumes that the Wilson coefficients 
 $(C_i)_{12}= (C_i)_{21}$, then the difference between ${\cal B}(B \to K \ell_1^- \ell_2^+ )$ and ${\cal B}(B \to K  \ell_2^- \ell_1^+)$ will be a measure of the interference term proportional to 
 $\mathrm{Re}[C_{9} C_S^*]$.

\subsection{$B\to K^\ast \ell_1 \ell_2$ and $B_s\to \phi \ell_1 \ell_2$}

These processes proceed via $B\to K^\ast (\to K\pi) \ell_1 \ell_2$ and $B_s\to \phi (\to K\bar K) \ell_1 \ell_2$. Since the expression for the angular distribution of 
the latter decay can be obtained by the trivial replacements in the expression for the former decay mode, we will focus on $\bar B\to \bar K^\ast (\to K^-\pi^+) \ell_1^- \ell_2^+$. 
As already stated in the previous subsection, we adopt the kinematics of ref.~\cite{Korner:1989qb} which is even more explicitly specified in ref.~\cite{Gratrex:2015hna} and fixed in such a way that they coincide with the conventions adopted in experiments at the LHC~\cite{Aaij:2015oid}. In the Appendix of 
the present paper we give necessary details concerning kinematics of this process.  Besides $\theta_\ell$ we also need $\theta_K$, the angle between 
the decay axis $-z$ and the direction of flight $K^-$ in the rest frame of $\bar K^\ast$ (cf. Fig.~\ref{fig:angles}). 
The angle between the planes spanned by $K\pi$ and $ \ell_1^- \ell_2^+$ respectively is denoted by $\phi$. 
In this case there are many more form factors parametrizing the hadronic matrix elements, namely, 
\begin{align}\label{def:FFV}
\langle \bar{K}^\ast(k)|\bar{s}\gamma^\mu(1-\gamma_5) b|\bar{B}(p)\rangle &= \varepsilon_{\mu\nu\rho\sigma}\varepsilon^{\ast\nu}p^\rho k^\sigma \frac{2 V(q^2)}{m_B+m_{K^\ast}}-i \varepsilon_\mu^\ast(m_B+m_{K^\ast})A_1(q^2)\\[.3em] 
&+i(p+k)_\mu (\varepsilon^\ast \cdot q)\frac{A_2(q^2)}{m_B+m_{K^\ast}}+i q_\mu(\varepsilon^\ast \cdot q) \frac{2 m_{K^\ast}}{q^2}[A_3(q^2)-A_0(q^2)],\nonumber\\[.7em] 
\langle \bar{K}^\ast(k)|\bar{s}\sigma_{\mu\nu} q^\nu(1-\gamma_5) b|\bar{B}(p)\rangle &= 2 i \varepsilon_{\mu\nu\rho\sigma} \varepsilon^{\ast\nu}p^\rho k^\sigma T_1(q^2)+[\varepsilon_\mu^\ast(m_B^2-m_{K^\ast}^2)-(\varepsilon^\ast \cdot q)(2p-q)_\mu]T_2(q^2)\nonumber\\[.3em] 
&+(\varepsilon^\ast \cdot q)\Big{[}q_\mu - \frac{q^2}{m_B^2-m_{K^\ast}^2}(p+k)_\mu \Big{]}T_3(q^2),
\end{align}
where $\varepsilon_\mu$ is the polarization vector of $K^\ast$, and the form factor $A_3(q^2)$ is not independent but related to $A_{1,2}(q^2)$ as, $2 m_{K^\ast} A_3(q^2)=(m_B+m_{K^\ast})A_1(q^2)-(m_B-m_{K^\ast})A_2(q^2)$.
The full angular distribution of the above decay reads~\footnote{Please notice that the convention used in eq.~(\ref{def:FFV}) is such that $\varepsilon_{0123}=+1$. }
\begin{equation}
\dfrac{\mathrm{d}^4 {\cal B} ({B}\to\bar{K}^{\ast}\to (K\pi) \ell_\alpha^-\ell_\beta^+)}{\mathrm{d}q^2\mathrm{d}\cos \theta_\ell \mathrm{d}\cos \theta_K \mathrm{d}\phi} = \dfrac{9}{32\pi}I(q^2,\theta_\ell,\theta_K,\phi),
\end{equation}
with
\begin{align}
I(q^2,\theta_\ell,\theta_K,\phi) = &I_1^s(q^2)\sin^2\theta_K + I_1^c(q^2)\cos^2\theta_K+[I_2^s(q^2)\sin^2\theta_K+I_2^c(q^2)\cos^2\theta_K]\cos 2\theta_\ell\nn\\[.4em] 
&+I_3(q^2)\sin^2\theta_K \sin^2\theta_\ell \cos 2\phi+I_4(q^2)\sin 2\theta_K \sin 2\theta_\ell \cos \phi \nn \\[.4em] 
&+ I_5(q^2) \sin 2\theta_K\sin \theta_\ell\cos\phi+[I_6^s(q^2)\sin^2\theta_K+I_6^c(q^2)\cos^2\theta_K]\cos \theta_\ell \nonumber \\[.4em] 
&+I_7(q^2)\sin 2\theta_K \sin \theta_\ell \sin \phi + I_8(q^2)\sin 2\theta_K \sin 2\theta_\ell \sin\phi \nonumber\\[.4em] 
&+I_9(q^2) \sin^2\theta_K \sin^2\theta_\ell \sin 2 \phi,
\end{align}
After integrating over angles the differential decay rate is simply 
\bea
{\mathrm{d}{\cal B}\over \mathrm{d}q^2}=\frac{1}{4}\left[3 I_1^c(q^2)+6 I_1^s(q^2)-I_2^c(q^2)-2I_2^s(q^2)\right]\ .
\eea 
The $q^2$-dependent angular coefficients are combinations of the decay's helicity amplitudes, which can also be expressed in terms of the transversity amplitudes $A_{\perp,\parallel,0,t}^{L(R)}\equiv A_{\perp,\parallel,0,t}^{L(R)}(q^2)$ as follows: 
\begin{align}
%\label{eq:helicityamplitudes}
A_{\perp}^{L(R)} &= {\cal N}_{K^\ast} \sqrt{2} \lambda_B^{1/2}\left[[(C_9+C_9')\mp(C_{10}+C_{10}')]\frac{V(q^2)}{m_B+m_{K^\ast}}+\frac{2 m_b}{q^2}(C_7+C_7') T_1(q^2)\right],\nn \\
%\label{eq:helicityamplitudespar}
A_{\parallel}^{L(R)} &=  -{\cal N}_{K^\ast} \sqrt{2}(m_B^2-m_{K^\ast}^2)\left[[(C_9-C_9')\mp (C_{10}-C_{10}')]\frac{A_1(q^2)}{m_B-m_{K^\ast}}+\frac{2 m_b}{q^2}(C_7-C_7')T_2(q^2)\right],\nn  
\end{align}
\begin{align}
A_0^{L(R)}&=-\frac{{\cal N}_{K^\ast}}{2 m_{K^\ast} \sqrt{q^2}}\Big{\lbrace} 2 m_b (C_7-C_7')\left[(m_B^2+3m_{K^\ast}^2-q^2)T_2(q^2)-\frac{ \lambda_B T_3(q^2)}{m_B^2-m_{K^\ast}^2} \right]\nn \\
&+[(C_9-C_9')\mp(C_{10}-C_{10}')]\cdot\left[ (m_B^2-m_{K^\ast}^2-q^2)(m_B+m_{K^\ast})A_1(q^2)-\frac{ \lambda_B A_2(q^2)}{m_B+m_{K^\ast}}\right] \Big{\rbrace} \nonumber\\[0.7 em]
\label{eq:helicityamplitudest}
A_{t}^{L(R)} &=  -{\cal N}_{K^\ast} \frac{\lambda_B^{1/2}}{\sqrt{q^2}}\left[(C_9-C_{9}') \mp (C_{10}-C_{10}') +\frac{q^2}{m_b+m_s}\left(\frac{C_S-C_S'}{m_1-m_2}\mp \frac{C_P-C_P'}{m_1+m_2}\right)\right] A_0(q^2),
\end{align}
where for shortness, $\lambda_B=\lambda(m_B,m_{K^\ast},\sqrt{q^2})$, $\lambda_q=\lambda(m_1,m_2,\sqrt{q^2})$, and 
\begin{equation}
{\cal N}_{K^\ast}=V_{tb}V_{ts}^\ast\left[ \frac{\tau_{B_d} G_F^2 \alpha^2}{3 \times 2^{10} \pi^5 m_B^3} \lambda_B^{1/2} \lambda_q^{1/2} \right]^{1/2}.
\end{equation}
The upper signs in the above formulas correspond to $A_i^L$ and the lower ones to $A_i^R$. 
Notice that $A_{t}$ also has the superscript $L(R)$, referring to the chirality of the lepton pair, which may appear unusual  when compared to the lepton flavor conserving case, and which we now explain.  
When $\ell_1=\ell_2$ the pseudoscalar density can be rewritten as
\begin{equation}
\bar{\ell} \gamma_5 \ell = \frac{q^\mu}{2 m_\ell} (\bar{\ell} \gamma_\mu \gamma_5 \ell),
\end{equation}
so that the contributions coming from the operator $\mathcal{O}_P^{(\prime)}$ can be absorbed in the amplitude $A_t$, which is associated to the timelike polarization vector of the virtual vector boson, $\epsilon_{V}^\mu(t)=q^\mu/\sqrt{q^2}$. 
A similar approach cannot be applied to the scalar operator $\mathcal{O}_S^{(\prime)}$, because $q^\mu (\bar{\ell} \gamma_\mu \ell) = 0$, and one must define a new amplitude $A_S$ to accommodate for  
the residual scalar contribution. In the LFV case, $m_1\neq m_2$, one can use the Ward identities to absorb both the scalar and the pseudoscalar densities in the vector and axial currents, respectively. 
Therefore, in the LFV case the amplitudes $A_t$ and $A_S$ are replaced by $A_t^{L(R)}$. Although the expressions for $A_t^{L,R}$ are ill-defined in the limit $m_1 = m_2$ we have checked that the angular coefficients are very well defined and one retrieves the standard formulas of ref.~\cite{Altmannshofer:2008dz}. 

Finally, in terms of the transversity amplitudes~(\ref{eq:helicityamplitudest}), the angular coefficients $I_{1-9}(q^2)$ are given by
\begin{align}
\label{eq:angular}
I_1^s(q^2) &=\biggl[|A_{\perp}^L|^2+|A_{\parallel}^L|^2+ (L\to R) \biggr]\frac{\lambda_q +2 [q^4-(m_1^2-m_2^2)^2]}{4 q^4}+\frac{4 m_1 m_2}{q^2}\mathrm{Re}\left(A_{\parallel}^L A_{\parallel}^{R\ast}+A_{\perp}^L A_{\perp}^{R\ast}\right), \nn \\
I_1^c(q^2) &= \bigl[|A_0^L|^2+|A_0^R|^2 \bigr]\frac{q^4-(m_1^2-m_2^2)^2}{q^4}+\frac{8 m_1 m_2}{q^2} \mathrm{Re}(A_0^L A_0^{R\ast}-A_t^L A_t^{R\ast}) \nn\\
&\hspace{3.5cm}-2\frac{(m_1^2-m_2^2)^2-q^2 (m_1^2+m_2^2)}{q^4}\bigl(|A_t^L|^2+|A_t^R|^2\bigr),\nonumber\\
I_2^s(q^2) &= \frac{\lambda_q}{4 q^4}[|A_\perp^L|^2+|A_\parallel^L|^2+(L\to R)], \nn  \\
I_2^c(q^2) &= - \frac{\lambda_q}{q^4}(|A_0^L|^2+|A_0^R|^2), \nn \\
I_3(q^2) &= \frac{\lambda_q}{2 q^4} [|A_\perp^L|^2-|A_\parallel^L|^2+(L\to R)],\nn \\
I_4(q^2) &= - \frac{\lambda_q}{\sqrt{2} q^4} \mathrm{Re}(A_\parallel^L A_0^{L\ast}+(L\to R)],\nn \\
I_5(q^2) &= \frac{\sqrt{2}\lambda_q^{1/2}}{q^2} \left[ \mathrm{Re}(A_0^L A_\perp^{L\ast}-(L\to R)) -\frac{m_1^2-m_2^2}{q^2} \mathrm{Re}(A_t^L A_\parallel^{L\ast}+(L\to R))\right], \nn \\
I_6^s(q^2) &=- \frac{2 \lambda_q^{1/2}}{q^2}[\mathrm{Re}(A_\parallel^L A_\perp^{L\ast}-(L\to R))],\nn  \\
I_6^c(q^2) &= - \frac{4\lambda_q^{1/2}}{q^2}\frac{m_1^2-m_2^2}{q^2} \mathrm{Re}(A_0^L A_t^{L\ast}+(L\to R)),\nn \\
I_7(q^2) &= - \frac{\sqrt{2}\lambda_q^{1/2}}{q^2} \left[ \mathrm{Im}(A_0^L A_\parallel^{L\ast}-(L\to R))+ \frac{m_1^2-m_2^2}{q^2} \mathrm{Im}(A_\perp^{L}A_t^{L\ast} +(L\to R))\right], \nn \\
I_8(q^2) &= \frac{\lambda_q}{\sqrt{2}q^4}\mathrm{Im}(A_0^{L}A_\perp^{L\ast} +(L\to R)), \nn \\
I_9(q^2) &=- \frac{\lambda_q}{q^4}\mathrm{Im}(A_\perp^L A_\parallel^{L\ast} +A_\perp^R A_\parallel^{R\ast}  ),
\end{align}
Once again, by taking the limit $m_1 \to m_2$, one retrieves the usual expressions for the coefficients of the angular distribution of $\bar B\to \bar K^\ast \ell^+\ell^-$. 
Our expressions agree with those recently presented in ref.~\cite{Gratrex:2015hna}, and are related to those given in ref.~\cite{Altmannshofer:2008dz} via $I_{4,6,7,9}\to - I_{4,6,7,9}$. In order to compare with the expressions for $A_t$ and $A_S$ from ref.~\cite{Altmannshofer:2008dz} one needs to identify 
\bea
A_t = \lim_{m_1\to m_2}\left(A_t^L-A_t^R \right),\qquad A_S = \lim_{m_1\to m_2}\left[ {m_1-m_2 \over \sqrt{q^2} }\left( A_t^L+A_t^R\right) \right]\,.
\eea

\subsection{Numerical significance}

To illustrate numerically the significance of the factors multiplying the Wilson coefficients, we use the form factors of ref.~\cite{Ball:2004rg} and distinguish the case of LFV arising from the vector operators, i.e. 
\begin{align}
\label{eq:numcoeffvec}
{\cal B}(\bar B\to \bar K^{(\ast)} \ell_1\ell_2) = 10^{-9} &\Big{(} a_{K^{(\ast)}}^{12} \Big{|}C_9+
C_9^{\prime}\Big{|}^2 + b_{K^{(\ast)}}^{12}  \Big{|}C_{10}+C_{10}^\prime \Big{|}^2\nonumber\\
&+ c_{K^{(\ast)}}^{12}  \Big{|}C_{9} -C_{9}^{\prime }\Big{|}^2+d_{K^{(\ast)}}^{12}  \Big{|}C_{10}-C_{10}^\prime \Big{|}^2\Big{)},
\end{align}
from the case in which the LFV comes from the scalar operators, 
\begin{align}
\label{eq:numcoeffscal}
{\cal B}(\bar B\to \bar K^{(\ast)} \ell_1\ell_2) = 10^{-9} &\Big{(} e_{K^{(\ast)}}^{12} \Big{|}C_S+
C_S^{\prime}\Big{|}^2 + f_{K^{(\ast)}}^{12}  \Big{|}C_{P}+C_{P}^\prime \Big{|}^2\nonumber\\
&+ g_{K^{(\ast)}}^{12}  \Big{|}C_{S} -C_{S}^{\prime }\Big{|}^2+h_{K^{(\ast)}}^{12}  \Big{|}C_{P}-C_{P}^\prime \Big{|}^2\Big{)},
\end{align}
The values of the factors multiplying the Wilson coefficients are obtained after integrating over all available $q^2$'s and are listed in Tab.~\ref{tab:1} and Tab.~\ref{tab:2}. 
\begin{table}[ht!]
\renewcommand{\arraystretch}{1.5}
\centering
\begin{tabular}{|c|cccc|cccc|}
\hline 
$\ell_1 \ell_2$ & $a_{K^{\ast}}^{12}$ & $b_{K^{\ast}}^{12}$ & $c_{K^{\ast}}^{12}$ & $d_{K^{\ast}}^{12}$  & $a_{K}^{12}$ & $b_{K}^{12}$ & $c_{K}^{12}$ & $d_{K}^{12}$ \\ \hline\hline
$e\mu$  & $7.8(9)  $ & $7.8(9) $ & $34(6) $ & $34(6)$ & $ 20(2) $ & $20(2)$ & $0$ & $0$\\  
$e\tau$  & $3.8(4) $ & $ 3.9(4) $ & $ 18(2)$ & $ 18(2) $ & $12.7(9) $ & $12.7(9) $ & $0$ & $0$\\  
$\mu\tau$  & $ 4.1 (5)$ & $ 3.6(4)$ & $ 18(2)$ & $17(2) $ & $12.5(1.0) $ & $12.9(9) $ & $0$ & $0$\\ \hline
\end{tabular}
\caption{\label{tab:1} \small \sl Values for the multiplicative factors defined in eq.~(\ref{eq:numcoeffvec}). The quoted uncertainties are at the $1\sigma$ level.}
\end{table}

\begin{figure}[t!]
\centering
\includegraphics[width=0.5\linewidth]{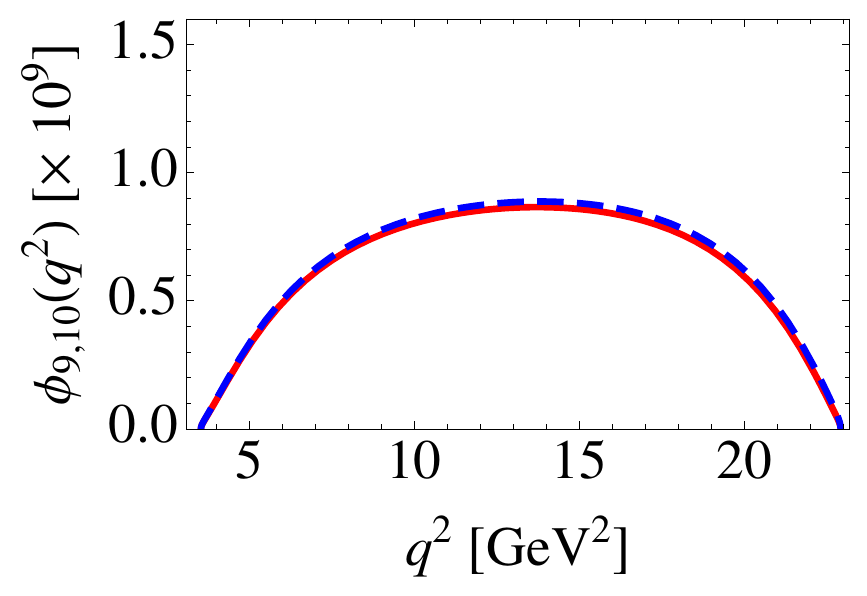}~\includegraphics[width=0.5\linewidth]{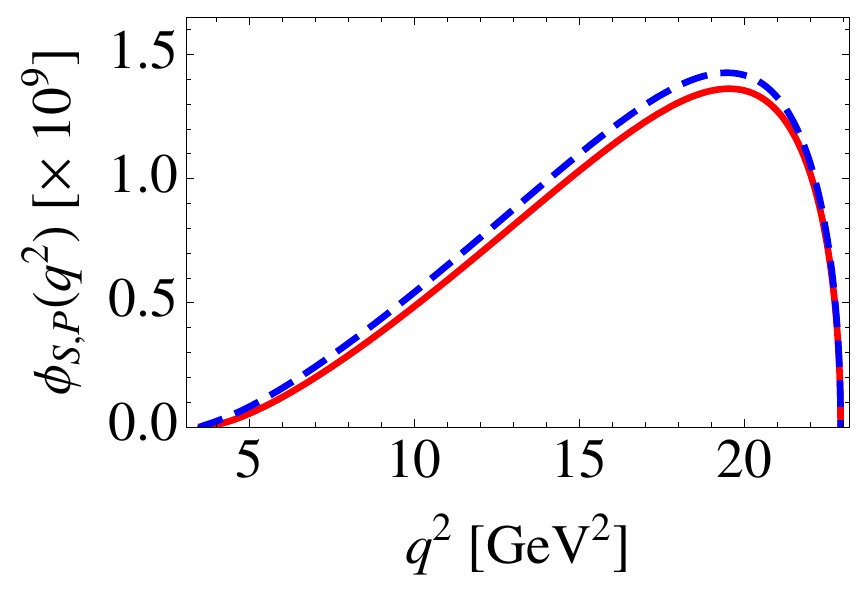}
\caption{\small \sl Coefficient functions $\phi_{9,10}(q^2) =\vert {\cal N}_{K}(q^2)\vert^2 \varphi_{9,10}(q^2)$ and $\phi_{S,P}(q^2) = \vert {\cal N}_{K}(q^2)\vert^2 \varphi_{S,P}(q^2)$ appearing in eq.~(\ref{eq:C910}), 
which after integration over $q^2$ give the factors $a_K^{\mu\tau}$, $b_K^{\mu\tau}$, $e_K^{\mu\tau}$ and $f_K^{\mu\tau}$ 
in eqs.~(\ref{eq:numcoeffvec},\ref{eq:numcoeffscal}). Full curves correspond to $\phi_{9}(q^2)$ and $\phi_{S}(q^2)$, while the dashed ones to $\phi_{10}(q^2)$ and $\phi_{P}(q^2)$.}
\label{fig:1}
\end{figure}

\begin{table}[ht!]
\renewcommand{\arraystretch}{1.5}
\centering
\begin{tabular}{|c|cccc|cccc|}
\hline 
$\ell_1 \ell_2$ & $e_{K^{\ast}}^{12}$ & $f_{K^{\ast}}^{12}$ & $g_{K^{\ast}}^{12}$ & $h_{K^{\ast}}^{12}$  & $e_{K}^{12}$ & $f_{K}^{12}$ & $g_{K}^{12}$ & $h_{K}^{12}$ \\ \hline\hline
$e\mu$  & $0$ & $0$ & $ 12(1)$ & $ 12(1)$ & $26.2(4)$ & $26.2(4)$ &$0$ & $0$ \\  
$e\tau$  &  $0$ & $0$ &$ 5.5(6)$ & $5.5(6)$ & $15.0(2)$ & $15.0(2) $  &$0$ & $0$\\  
$\mu\tau$  &  $0$ & $0$ & $5.2(6)$ & $5.8(7)$ & $14.4(2) $ & $15.5(2)$  &$0$ & $0$\\ \hline
\end{tabular}
\caption{\label{tab:2} \small \sl Values for the multiplicative factors defined in eq.~(\ref{eq:numcoeffscal}) to $1\sigma$ accuracy.}
\end{table}
Notice also that the functions which are being integrated to obtain those factors have a peculiar feature: those which multiply $|C_{9,10}  \pm 
C_{9,10}^\prime |^2$  are more pronounced in the intermediate $q^2$ region, whereas those multiplying $|C_{S,P}  \pm 
C_{S,P}^\prime |^2$ are mostly receiving contributions from the large $q^2$ region. To illustrate this feature, we show in Fig.~\ref{fig:1} the coefficient functions $\varphi_{9,10}(q^2)$ [$\varphi_{S,P}(q^2)$], which upon integration amount to $a_{K}^{\mu\tau}$ and $b_K^{\mu\tau}$ [$e_{K}^{\mu\tau}$ and $f_K^{\mu\tau}$].~\footnote{The purpose of the plots shown in Fig.~\ref{fig:1} is to illustrate the shapes of $\phi_{i}(q^2) =\vert {\cal N}_{K}(q^2)\vert^2 \varphi_{i}(q^2)$ and the uncertainties on hadronic form factors were omitted in the plots. Those uncertainties, instead, have been properly accounted for when computing the factors listed in Tab.~\ref{tab:2}.}
%%%%%%%%%%%%%%%%%%%%%%%%%%%%%%%%%%%%%%%%
%%%%%%%%%%%%%%%%%%%%%%%%%%%%%%%%%%%%%%%%
%%%%%%%%%%%%%%%%%%%%%%%%%%%%%%%%%%%%%%%%
%%%%%%%%%%%%%%%%%%%%%%%%%%%%%%%%%%%%%%%%

Furthermore in the case of LFV generated by the scalar operators the lifted helicity suppression of the leptonic decay~(\ref{Bsformula}) leads to the following hierarchy among different modes:
\begin{equation}
\label{eq:ineqS}
C_{S,P}^{(\prime )}\neq 0, C_{9,10}^{(\prime )}=0:\qquad \mathcal{B}(B_s\to \ell_1\ell_2)>\mathcal{B}(B\to K  \ell_1\ell_2)>\mathcal{B}(B\to K^\ast  \ell_1\ell_2).
\end{equation}
That hierarchy is inverted for the LFV processes generated by the vector operators, namely
\begin{equation}
\label{eq:ineqV}
C_{S,P}^{(\prime )}= 0, C_{9,10}^{(\prime )}\neq 0:\qquad \mathcal{B}(B_s\to  \ell_1\ell_2)<\mathcal{B}(B\to K  \ell_1\ell_2)<\mathcal{B}(B\to K^\ast  \ell_1\ell_2).
\end{equation}
Of course the above discussion is valid as long as we do not consider the case of LFV generated by both the scalar and vector operators, which we will not discuss in what follows anyway.

\section{ \label{sec:3} A case of $C_{S,P}\neq 0$: Coupling to Higgs}

In this section we focus on the specific example of a scenario in which the LFV is generated through the scalar operators. We will relate the $2.2\sigma$ excess 
of $h\to\mu\tau$ observed by CMS~\cite{Khachatryan:2015kon}, to the decays $B_s \to \mu \tau$ 
and $B\to K^{(\ast)}\mu\tau$.~\footnote{Please note that Atlas too observed an excess of  $h\to\mu\tau$, although the significance is only at the $1.2\sigma$ level~\cite{Aad:2015gha}. They reported ${\cal B}(h\to\mu\tau)_{\rm Atlas} = 0.77(62)\%$, 
to be compared with ${\cal B}(h\to\mu\tau)_{\rm CMS} = 0.84^{+0.39}_{-0.37}\%$.} 

\begin{figure}[t!]
\hspace*{-7mm}\includegraphics[width=0.53\linewidth]{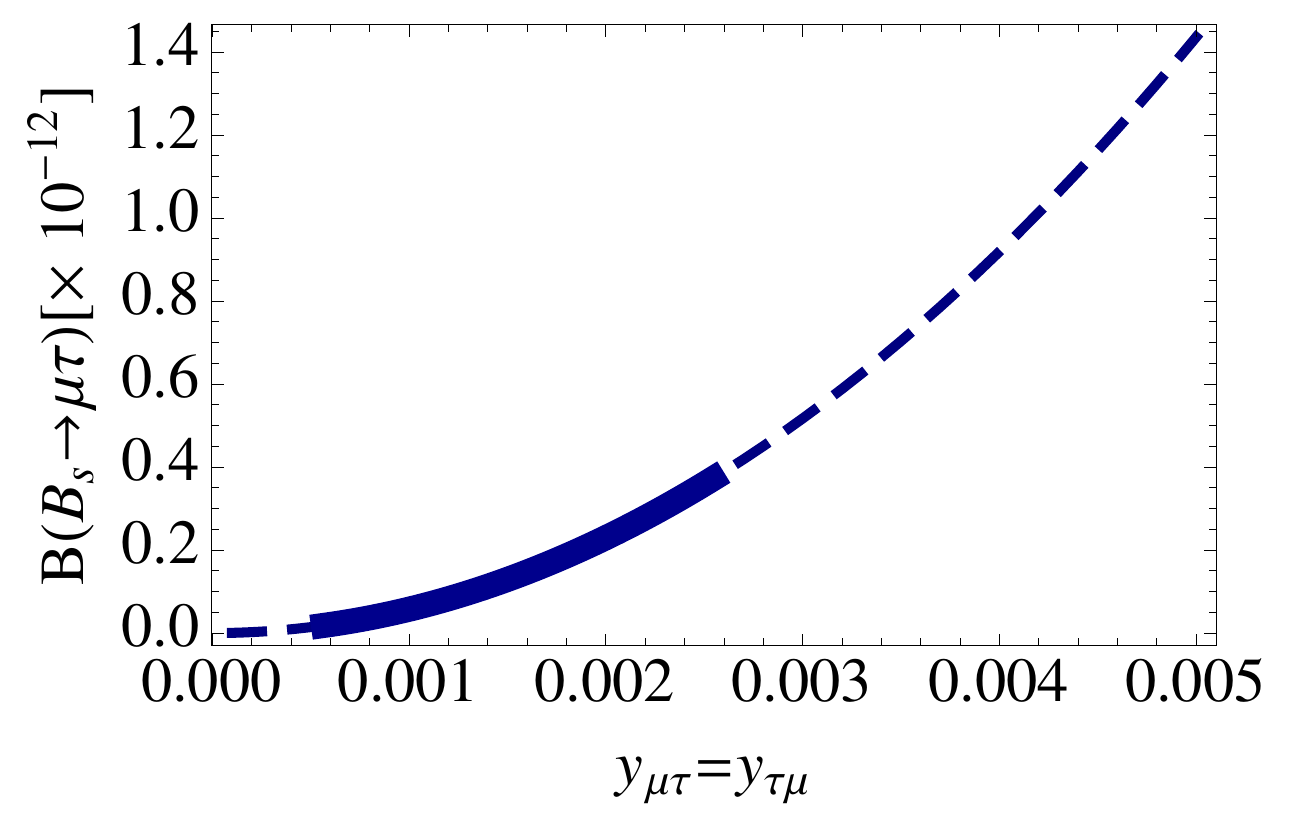}~\includegraphics[width=0.55\linewidth]{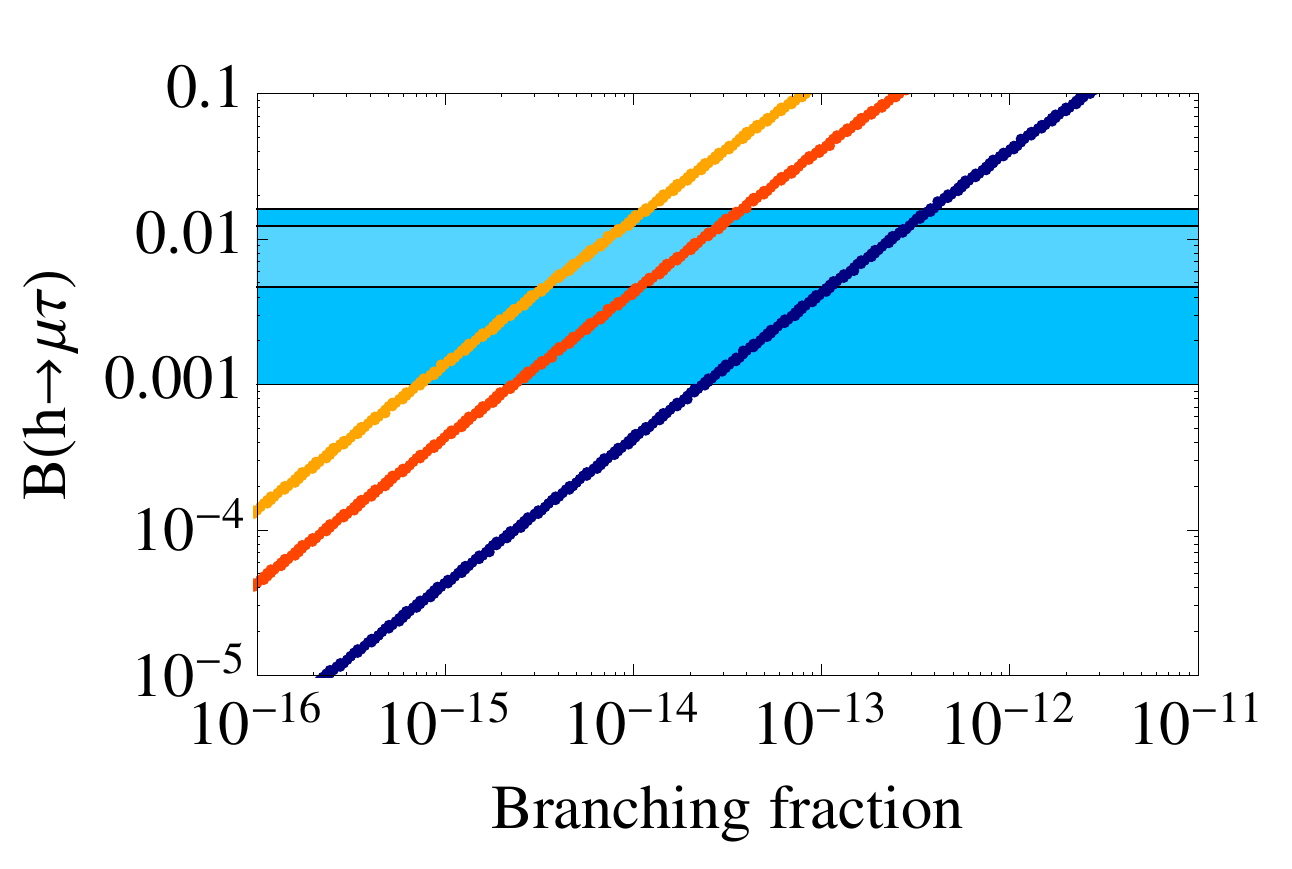}
\caption{\small \sl In the left panel we plot the branching fraction of $B_s\to\tau\mu$ decay as a function of $y_{\tau\mu}=y_{\mu\tau}$. The segment shown in full line is the one corresponding to $y_{\tau\mu}$ extracted from ${\cal B}(h\to \tau \mu)$ to $2\sigma$. In the right panel we show ${\cal B}(h\to\tau\mu)$ by the brighter (darker) stripe to $1 (2) \sigma$ as reported by CMS, versus ${\cal B}(B_s\to\tau\mu)$ [blue], ${\cal B}(B\to K\tau\mu)$ [red], ${\cal B}(B\to K^\ast \tau\mu)$ [orange].}
\label{fig:2}
\end{figure}

In the scenarios in which the physics BSM comes solely from the modification of the Higgs sector, the decay $h\to\mu\tau$ can be described by the Yukawa Lagrangian,
\begin{align}
	\mathcal{L}_Y^{\mathrm{eff}} = - y_{ij} \bar{\ell}^i_L \ell^j_R h + \mathrm{h.c.}
\end{align}
The non-diagonal couplings $y_{ij}$ can originate in the mixing of the Higgs doublet with additional scalar doublets, and the above Lagrangian is fully adequate if the masses of other Higgs states are larger than $m_h$~\cite{Harnik:2012pb}. 
The only Wilson coefficients in eq.~(\ref{heff}) that receive non-negligible contributions through the scalar penguin diagrams are~\cite{Dedes:2003kp}
\begin{align}\label{eq:csp}
C_{S,P} = - \frac{y_{\mu\tau}\pm y_{\tau\mu}^\ast}{2}&\frac{m_b v}{16 m_W^2\sin^2\theta_W} \times \nn\\
& \left(  \frac{6 x_t}{x_h} 
- \frac{2 x_t^3}{  (1-x_t)^3}\ln x_t +\frac{4 x_t^2}{  (1-x_t)^3}\ln x_t - \frac{x_t^2}{ (1-x_t)^2} + \frac{3 x_t}{ (1-x_t)^2}  
\right), 
\end{align}
where $x_{t,h}= m_{t,h}^2/m_W^2$, $v= 246\ \gev$, and the upper (lower) signs corresponds to $C_S$ ($C_P$).~\footnote{We emphasize that in the situation in which $y_{ij}$ arise as a loop effect, the Wilson coefficients in Eq.~(\ref{eq:csp}) would clearly be incomplete.} 
Using the CMS result, ${\cal B}(h\to \mu\tau ) = 0.84^{+0.39}_{-0.37}\%$, and the formula $\Gamma(h\to \mu\tau )= \left(|y_{\mu\tau}|^2+|y_{\tau\mu}|^2 \right)m_h/(8\pi)$, one obtains
\begin{equation}
\label{eq:limitsony}
1.9 [0.8] < 10^3 \times \sqrt{|y_{\mu\tau}|^2+|y_{\tau\mu}|^2 }< 3.2 [3.6]\qquad\text{at }\,68\%\, [95\%]\,\text{CL},
\end{equation}
which then amounts to 
\begin{equation}
8.4 [3.5]<10^4\times \sqrt{|C_S|^2+|C_P|^2}<  14.2 [16.0]\qquad\text{at }\,68\%\, [95\%]\,\text{CL}.
\end{equation}
Notice that the couplings $y_{\mu \tau}$ and $y_{\tau\mu }$ are tacitly assumed to be complex, in which case the quantity $|y_{\mu\tau}|^2+|y_{\tau\mu}|^2$ is not enough to completely determine 
the decay amplitudes of the processes described here. One possibility to tackle this issue is to use eqs.~(\ref{Bsformula}) and (\ref{eq:csp}), write
\begin{equation}
\mathcal{B}(B_s \to \mu \tau ) \propto (m_{B_s}^2-m_\mu^2-m_\tau^2)(|y_{\mu\tau}|^2+|y_{\tau\mu}|^2)-2m_\mu m_\tau \mathrm{Re}(y_{\mu\tau}
 y_{\tau\mu}^\ast ),
\end{equation}
and combine it with the constraint coming from $\mathcal{B}(\tau \to \mu \gamma )$, namely~\footnote{ For the Wilson coefficient we take the result of ref.~\cite{Harnik:2012pb}, 
\bea
C_L^\gamma = {1\over 12 m_h^2}\frac{m_\tau}{v} y_{\tau\mu}^\ast \left( - 4 + 3 \ln\frac{m_h^2}{m_\tau^2}\right) + 0.055 {y_{\tau\mu}^\ast  \over (125\ \gev)^2},
\nn\eea
and $C_R= \left. C_L^\ast\right|_{\mu\leftrightarrow \tau}$.}
\bea
\mathcal{B}(\tau \to \mu \gamma )  = {\alpha m_\tau^5\over  64 \pi^4 \Gamma_\tau} \left( |C_L^\gamma|^2+|C_R^\gamma|^2\right), 
\eea
and $\mathcal{B}(\tau \to \mu \gamma )_{\rm exp.}<4.4\times 10{-8}$~\cite{PDG}. As of now nothing can be said about the complex phases of these couplings and in what follows we assume them to be zero. 
As indicated in eq.~(\ref{eq:ineqS}) the most sensitive channel to the presence of $C_S\neq 0$ is the leptonic decay mode ${\cal B}(B_s\to \ell_1\ell_2 )$. To exacerbate the phenomenon we focus on the $\mu\tau$-decay channel and show in Fig.~\ref{fig:2} its dependence on the coupling $y_{\mu\tau}=y_{\tau\mu}$. We also show the plot of ${\cal B}(h\to \tau \mu)$ versus the branching fractions of the modes we are interested in for increasing values of $y_{\tau\mu}$. The horizontal stripe correspond to the $1\sigma$ (darker) and $2\sigma$ (brighter) result reported by CMS.
%%%%%%%%%%%%%%%%%%%%%%%%%%%%%%%%%%%%%%%%
%%%%%%%%%%%%%%%%%%%%%%%%%%%%%%%%%%%%%%%%
%%%%%%%%%%%%%%%%%%%%%%%%%%%%%%%%%%%%%%%%
%%%%%%%%%%%%%%%%%%%%%%%%%%%%%%%%%%%%%%%%

Finally, the bounds on the LFV modes obtained in this way are:
\bea
&&{\cal B}(B_s \to \mu \tau ) <3.9\times 10^{-13}\,,\nn\\
&&{\cal B}(B \to K \mu \tau )< 3.8 \times 10^{-14}\,,\nn\\
&&{\cal B}(B \to K \mu \tau ) < 1.2 \times 10^{-14}\,.
\eea
These bounds are too small for current experimental searches. The purpose of this section, however, was only to illustrate the effect of LFV generated through the scalar couplings extracted from the experimental bound on ${\cal B}(h\to\mu\tau)$. 
If the origin of such a coupling is different from the one discussed here, the above bounds could be larger (less stringent) but the hierarchy given in eq.~(\ref{eq:ineqS}) will still hold true.

%%%%%%%%%%%%%%%%%%%%%%%%%%%%%%%%%%%%%%%%
%%%%%%%%%%%%%%%%%%%%%%%%%%%%%%%%%%%%%%%%
%%%%%%%%%%%%%%%%%%%%%%%%%%%%%%%%%%%%%%%%
%%%%%%%%%%%%%%%%%%%%%%%%%%%%%%%%%%%%%%%%
%%%%%%%%%%%%%%%%%%%%%%%%%%%%%%%%%%%%%%%%
%%%%%%%%%%%%%%%%%%%%%%%%%%%%%%%%%%%%%%%%

\section{ \label{sec:4} A case of $C_{9,10}\neq 0$: Coupling to $Z^\prime$}

In this section we revisit a $Z^\prime$-model, already discussed in the context of this problem in ref.~\cite{Crivellin:2015era}. It illustrates the case in which the LFV is generated by the (axial-)vector operators. 
Furthermore, since our bounds somewhat differ from those reported in ref.~\cite{Crivellin:2015era} we believe it is worth discussing it in more detail. 
The most general lagrangian involving $Z^\prime$ reads
\begin{equation}
\mathcal{L}_{Z^\prime} \supset g_{\ell_i \ell_j}^L \bar{\ell_i }\gamma^\mu P_L \ell_j \ Z_\mu^\prime + g_{sb}^L \bar{s}\gamma^\mu P_L b \ Z_\mu^\prime + (L\to R),
\end{equation}
where we assume that the $Z^\prime$ boson couples only to the second and third generations of quarks and leptons. Since the scale of new physics is assumed to be well 
above the electroweak one, the $SU(2)_L$ gauge invariance has to be preserved, which then implies that, for example, $g_{\ell_i \ell_j}^L=g_{\nu_i\nu_j}^L$ and $g_{sb}^L=g_{ct}^L$.

After integrating out the $Z^\prime$, the relevant Wilson coefficients read
\begin{align}\label{eq:c910}
C_9^{(\prime)\mu\tau} &= -\frac{\pi}{\sqrt{2}m_{Z'}^2}\frac{1}{\alpha G_F V_{tb} V_{ts}^\ast} g_{sb}^{L(R)}(g_{\mu\tau}^R+g_{\mu\tau}^L), \nn\\
C_{10}^{(\prime)\mu\tau} &= -\frac{\pi}{\sqrt{2}m_{Z'}^2}\frac{1}{\alpha G_F V_{tb} V_{ts}^\ast} g_{sb}^{L(R)}(g_{\mu\tau}^R-g_{\mu\tau}^L),
\end{align}
where the primed Wilson coefficients are proportional to $g_{sb}^{R}$.  To get the value of $g_{sb}^{L(R)}$ we use the information on the $B_s-\overline B_s$ mixing amplitude and add the contribution coming from the couplings to $Z^\prime$. 
More specifically, we add
\bea
{\cal H}_{\rm eff}^{Z^\prime} = C_1 (\bar b \gamma_\mu P_L s) (\bar b \gamma^\mu P_L s) + C_1^\prime (\bar b \gamma_\mu P_R s) (\bar b \gamma^\mu P_R s) + C_5 (\bar b_i P_L s_j) (\bar b_j P_R s_i) , 
\eea
to the SM contribution. Wilson coefficients are easily computed at $\mu \approx m_{Z^\prime}$ and read,
\bea
C_1^{(\prime )} = { \left( g_{sb}^{L(R)} \right)^2   \over 2 m_{Z^\prime}^2}\,, \quad C_5 = -  {2  g_{sb}^{L}g_{sb}^R \over m_{Z^\prime}^2}\,, 
\eea
which then, combined with 
\bea
&&\langle \bar B_s^0\vert \bar b \gamma_\mu (1-\gamma_5) s\, \bar b \gamma^\mu (1-\gamma_5) s \vert B_s^0\rangle = \frac{8}{3} f_{B_s}^2 m_{B_s}^2 B_1(\mu)\,,\nn\\
&&\langle \bar B_s^0\vert \bar b_i  (1-\gamma_5) s_j\, \bar b_j  (1-\gamma_5) s \vert B_s^0\rangle = \frac{2}{3} f_{B_s}^2 m_{B_s}^2 \left( \frac{m_{B_s}}{m_b(\mu)+m_s(\mu)} \right)^2 B_5(\mu)\,,
\eea
lead to 
\begin{align}\label{eq:XX}
{\Delta m_{B_s}^{\rm exp.} \over \Delta m_{B_s}^{\rm SM}} =  1 + {2\pi^2 \over G_F^2 m_W^2 |V_{tb}V_{ts}^\ast |^2 \eta_B S_0(x_t) m_{Z^\prime}^2} \left[ \eta_1 (g_{sb}^{L})^2 + \eta_1(g_{sb}^R)^2 - \eta_5{B_5(m_b)\over B_1(m_b)} \left( {m_{B_s}\over m_b + m_s}
\right)^2 g_{sb}^{L}g_{sb}^R 
\right] ,
\end{align}
where $\eta_{1,5}$ account for the evolution of the Wilson coefficients from the scale $\mu = m_{Z^\prime}$ down to $\mu=m_b$, which we evaluate using the two-loop QCD anomalous dimensions to find~\cite{Buras:2000if,Buras:2001ra}  
\bea
&&\eta_1= 0.79 [0.80], \quad  \eta_5= 0.89 [0.90]\quad\mathrm{for}\; m_{Z^\prime}=1\ \tev,\nn\\
&&\eta_1= 0.77 [0.78], \quad  \eta_5= 0.88 [0.89]\quad\mathrm{for}\; m_{Z^\prime}=2\ \tev,
\eea
where in the square brackets we quote the values obtained to leading order in QCD. The hadronic quantities entering eq.~(\ref{eq:XX}) have been computed by means of numerical simulations of QCD on the lattice in ref.~\cite{Carrasco:2013zta} and read,  
\bea
f_{B_s}= 228(8)\ \mev,\qquad B_1^\msbar(m_b) = 0.86(3),\qquad 
B_5^\msbar(m_b) = 1.57(11)\,.
\eea
Since we consider here only the scenarios in which either $g_{sb}^{L}\neq 0$, $g_{sb}^{R}=0$, or  $g_{sb}^{R}\neq 0$, $g_{sb}^{L}=0$, the last term in eq.~(\ref{eq:XX}) will always be zero for us. Therefore, keeping in mind that $({\Delta m_{B_s}^{\rm exp.}/\Delta m_{B_s}^{\rm SM}})= 1.02(10)$, and by using the above ingredients we find, to $2\sigma$ accuracy,
\bea
{\vert g_{sb}^{L(R)}\vert \over m_{Z^\prime}}\leq 1.6(8) \times 10^{-3}\ {\rm TeV}^{-1}\ .
\eea
Another coupling needed in eq.~(\ref{eq:c910}) is $g_{\mu\tau}^L$ which can be extracted from the deviation of the measured ${\cal B}(\tau \to \mu \bar \nu_\mu \nu_\tau)_{\rm exp.}= 17.33(5)\%$~\cite{PDG} with respect to its Standard Model prediction ${\cal B}(\tau \to \mu \bar \nu_\mu \nu_\tau)_{\rm theo.}^{\rm SM}= 17.29(3)\%$, namely~\cite{Crivellin:2015era,Altmannshofer:2014cfa}:~\footnote{For completeness we remind the reader that the SM expression for the leptonic decay rate reads:
\bea
{\cal B}(\tau \to \mu \bar \nu_\mu \nu_\tau) =\tau_\tau {G_F^2 m_\tau^5\over 192\pi^3 } {\cal E}\left[1- 8 {m_\mu^2\over m_\tau^2}+ 8 \left( {m_\mu^2\over m_\tau^2} \right)^3- \left( \frac{m_\mu^2}{m_\tau^2}\right)^4
-12  \left( \frac{m_\mu^2}{m_\tau^2}\right)^2\ln \left( \frac{m_\mu^2}{m_\tau^2}\right)\right],
\nn \eea 
where the radiative and the correction due to the propagation of $W$ amounts to ${\cal E}= 0.996$.  } 
\begin{align}
\delta {\cal B}_{\tau\mu}= {\cal B}(\tau \to \mu \bar \nu_\mu \nu_\tau)_{\rm exp.}-{\cal B}(\tau \to \mu \bar \nu_\mu \nu_\tau)_{\rm theo.}^{\rm SM} =-{m_\tau^5\over 1536 \pi^3 \Gamma_\tau  m_{Z^\prime}^2} {8 G_F\over \sqrt{2}}  \left( g_{\mu\tau}^L\right)^2 +{\mathcal O}(1/m_{Z^\prime}^4).
\end{align}
Finally, the last coupling needed in eq.~(\ref{eq:c910}) is $g_{\mu\tau}^R$, which can be bounded from ${\cal B}(\tau \to \mu \mu \mu)_{\rm exp.}  < 2.1\times 10^{-8}$~\cite{PDG}, by using the expression~\cite{Crivellin:2015era}
\bea\label{eq:3mu}
{\cal B}(\tau \to 3 \mu) = {m_\tau^5\over 1536 \pi^3 \Gamma_\tau  m_{Z^\prime}^4} \left( g_{\mu\mu}^L\right)^2 \left[ 
2 \left( g_{\mu\tau}^L\right)^2 + \left( g_{\mu\tau}^R\right)^2 \right].
\eea
Besides $g_{\mu\tau}^L$ which we discussed above, we need the value of $g_{\mu\mu}^L$, which can be obtained from a fit to the $b\to s\mu\mu$ data. To that end we consider two scenarios: the one in which the new physics contribution to the lepton flavor conserving channel comes entirely from $g_{sb}^L$, i.e. $C_{9}^{\mu\mu} = -C_{10}^{\mu\mu}$, and the case in which the coupling to quarks is entirely right-handed,  $g_{sb}^R$, and the Wilson coefficients satisfy  $C_{9}^{\prime \mu\mu} = -C_{10}^{\prime \mu\mu}$. 
Concerning the value of $C_9^{(\prime)}$ we can derive it as in ref.~\cite{Becirevic:2015asa}, by relying on the safest quantities as far as hadronic uncertainties are concerned, which to $2\sigma$-accuracy results in
\begin{align}\label{eq:ourC910}
&C_9^{\mu\mu}\in [-0.52,-0.19],\qquad C_9^{\mu\mu\ \prime} \in [-0.41,-0.08],
\end{align}
and makes $R_K$ consistent with experiment.~\footnote{One can otherwise use the results of the global fit: $C_9^{\mu\mu}\in [-1.04,-0.34]$,
and $C_9^{\mu\mu\ \prime}  \in [-0.05,0.43]$ from ref.~\cite{Descotes-Genon:2015uva}, or $C_9^{\mu\mu}\in [-0.91,-0.18]$,
and $C_9^{\mu\mu\ \prime}  \in [-0.12,0.33]$ from ref.~\cite{Altmannshofer:2014rta}, also valid to $2\sigma$-accuracy.
}
%%%%%%%%%%%%%%%%%%%%%%%%%%%%%%%%%%%%%%%%
%%%%%%%%%%%%%%%%%%%%%%%%%%%%%%%%%%%%%%%%
\begin{figure}[t!]
\centering
\hspace*{-7mm}\includegraphics[width=0.53\linewidth]{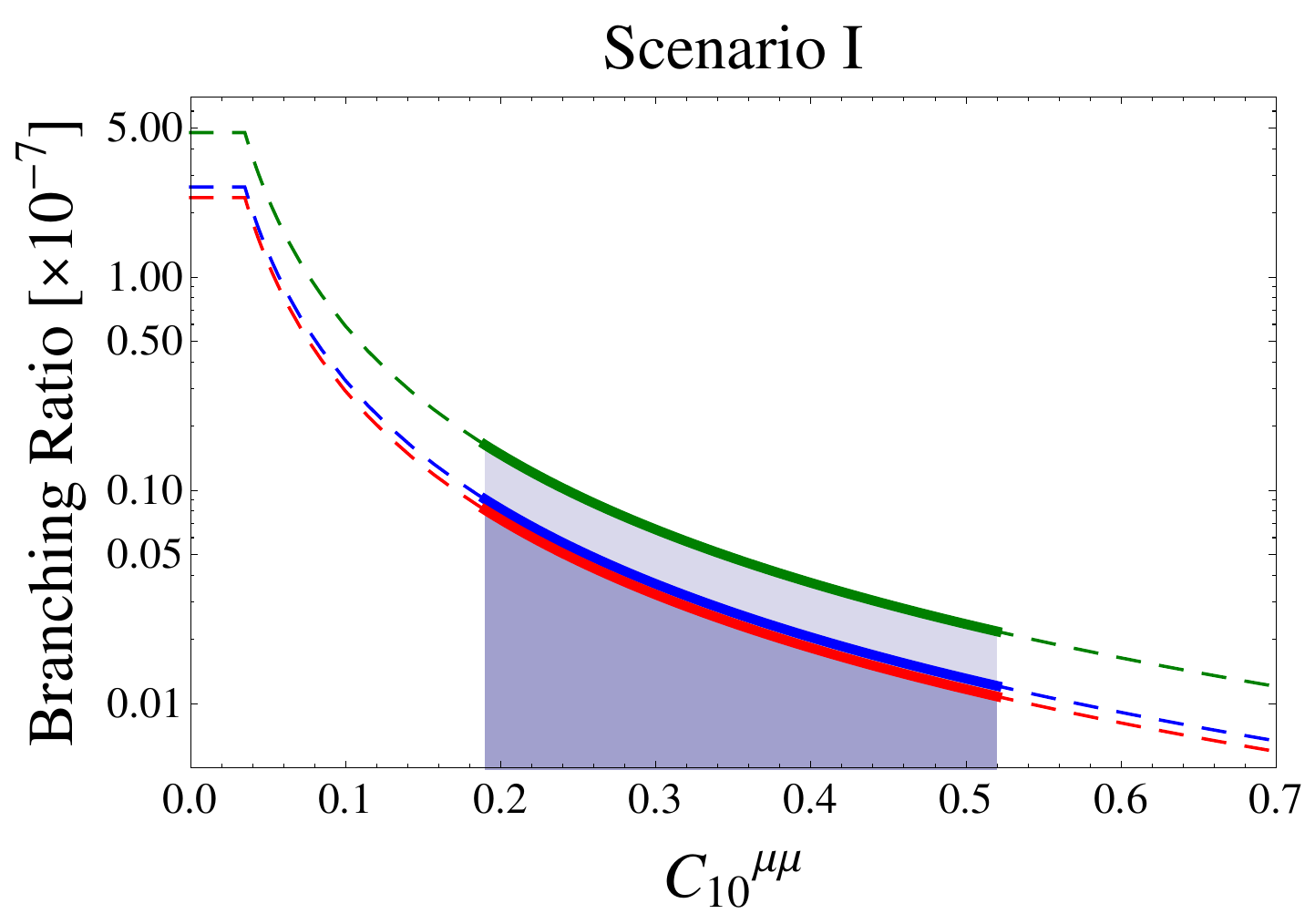}~\includegraphics[width=0.53\linewidth]{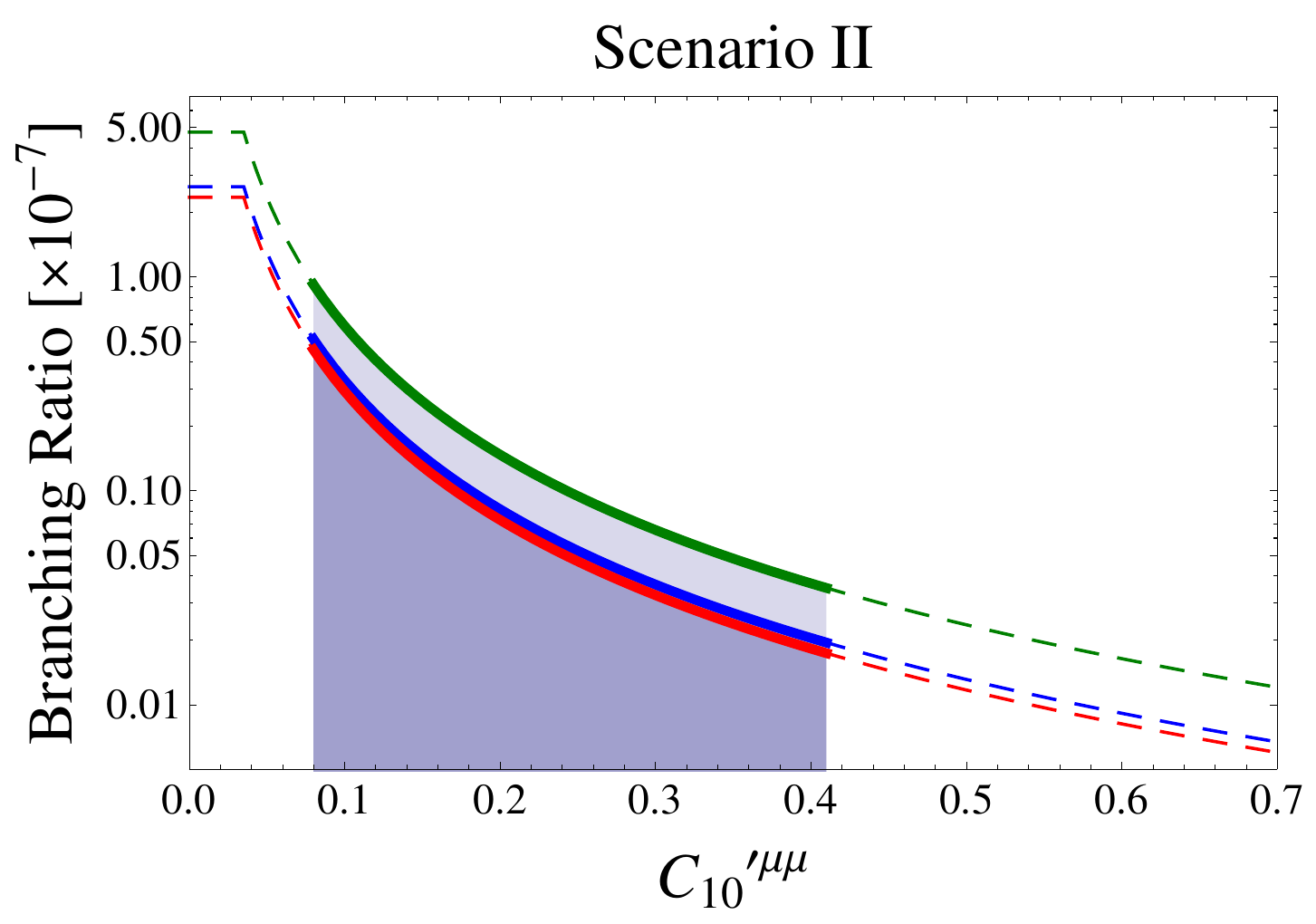}
\caption{\small \sl Upper bounds on the branching fractions $\mathcal{B}(B\to K^\ast  \mu\tau) > \mathcal{B}(B\to K  \mu\tau)>\mathcal{B}(B_s\to  \mu\tau)$, as a function of the BSM contribution to $C_{10}^{(\prime )\mu\mu}$ extracted from the LFC decay modes in two setups: in Scenario~I 
we use $C_{10}^{\mu\mu}=-C_{9}^{\mu\mu}$ while keeping $C_{9,10}^{\prime\ \mu\mu}=0$, and in Scenario~II we take $C_{10}^{\prime\ \mu\mu}=-C_{9}^{\prime\ \mu\mu}$ with $C_{9,10}^{\mu\mu}=0$. Shaded regions correspond to the values given in eq.~(\ref{eq:ourC910}), obtained by combining ${\cal B}(B_s\to \mu\mu)$ with the high $q^2$ bin of $d{\cal B}(B\to K \mu\mu)/dq^2$, which result in $R_K$ consistent with experiment, cf. ref.~\cite{Becirevic:2015asa}. See text for discussion concerning the couplings $g_{sb,\mu\tau}^{L(R)}$.}
\label{figX}
\end{figure}
%%%%%%%%%%%%%%%%%%%%%%%%%%%%%%%%%%%%%%%%
%%%%%%%%%%%%%%%%%%%%%%%%%%%%%%%%%%%%%%%%
Such an obtained $g_{\mu\mu}^L$ is then used to get $g_{\mu\tau}^R$ by means of eq.~(\ref{eq:3mu}). Notice, however, that for very small values of $g_{\mu\mu}^L$ the value of $g_{\mu\tau}^R$ can be excessively large if we require the saturation of the experimental bound. In those cases we invoke the perturbativity requirement and set the bound to $|g_{\mu\tau}^R| \leq 1$. 
With all above ingredients in hands we can compute $C_{9,10}^{\mu\tau (\prime )}$ by means of eq.~(\ref{eq:c910}), and then use the obtained values to predict the upper bounds for the rates of the decay modes we discuss here. 
In Fig.~\ref{figX} we show such bounds for both scenarios: (i) in Scenario~I we use $C_9^{\mu\mu}= - C_{10}^{\mu\mu}$ to determine $g_{\mu\mu}^L$, while in (ii) Scenario~II we use the condition $C_9^{\prime\ \mu\mu}= - C_{10}^{\prime\ \mu\mu}$. The resulting bounds satisfy the hierarchy noted in eq.~(\ref{eq:ineqV}). We focus on the values of $C_9^{(\prime ) \mu\mu}= - C_{10}^{(\prime ) \mu\mu} \neq 0$ (and $C_9^{(\prime ) ee}=0$) which give $R_K$ consistent with the one measured at LHCb. That range of values correspond to the shaded regions in the plots in Fig.~\ref{figX}. The resulting bounds are:
\begin{center}\renewcommand{\arraystretch}{1.8}
\begin{tabular}[H]{ccc}   
 \hline
Scenario & I & II \\ \hline
${\cal B}(B \to K^\ast \mu \tau ) \leq$ & $1.6 \times 10^{-8}$ & $9.3 \times 10^{-8}$ \\
${\cal B}(B \to K \mu \tau ) \leq$ & $0.9 \times 10^{-8}$ & $5.2 \times 10^{-8}$ \\
${\cal B}(B_s \to \mu \tau ) \leq$  & $0.8 \times 10^{-8}$ & $4.6 \times 10^{-8}$ \\
 \hline
\end{tabular}
\end{center}
We stress once again that the above bounds are obtained after assuming that the BSM physics effects come in the scenarios with either $C_9=-C_{10}$, or $C_9^\prime=-C_{10}^\prime$. In other words either $g_{sb}^R=0$, or $g_{sb}^L=0$. If no assumption about the BSM physics is being made, and both $g_{sb}^{L}$ and $g_{sb}^{R}$ were left free, then the third term in the brackets of Eq.~(\ref{eq:XX}) would play an important role and the resulting bounds on the above decay modes would be weaker. 

\section{\label{sec:5} Summary}
In the present paper we discussed the possibility of observing the LFV modes in exclusive decays based on $b\to s\ell_1^\pm \ell_2^\mp$. Starting from the low energy effective hamiltonian, we derived the 
expressions for decay rates for $B_s\to \ell_1\ell_2$,  $B\to K \ell_1\ell_2$, $B\to K^\ast (\to K\pi ) \ell_1\ell_2$ and similar modes. We show that the extra contributions proportional to the difference between lepton masses arise in the case of LFV modes, 
thus requiring particular care when trying to average the (lepton) charge-conjugated modes.
We then examined the situation in which the LFV is generated by the (pseudo-)scalar operators, to distinguish it from the one in which the LFV comes from the coupling to (axial-)vector operators.  In the former case we find that most of the events would occur at larger values of $q^2$, while in the latter case the events are expected to be equidistributed over a large window of $q^2$'s. Furthermore, we find that the hierarchy of the branching fractions of our modes change: while in the case of coupling to the (axial-)vector operators we find $\mathcal{B}(B_s\to  \ell_1\ell_2)<\mathcal{B}(B\to K  \ell_1\ell_2)<\mathcal{B}(B\to K^\ast  \ell_1\ell_2)$, in the case of coupling to the (pseudo-)scalar operators we get  $\mathcal{B}(B_s\to  \ell_1\ell_2)>\mathcal{B}(B\to K  \ell_1\ell_2)>\mathcal{B}(B\to K^\ast  \ell_1\ell_2)$. To illustrate both cases we first used a phenomenological Lagrangian that encodes ${\cal B}(h\to\mu\tau)\neq 0$, as recently suggested by CMS, to derive the bounds that seem to be too low for these decay modes to be probed experimentally. In the second case we revisited a $Z^\prime$ model in which a (small) tree level flavor changing neutral couplings are allowed, and after a short discussion concerning the specific scenarios and the channels allowing to bound the relevant LFV couplings, we derive bounds which generically suggest the branching fractions of all the modes we consider to be less than a few times $10^{-8}$, which are thus more likely to be probed experimentally.

As a by-result of our analysis we revisited the computation of the angular distribution of $B\to K^\ast (\to K\pi ) \ell_1\ell_2$, which is often a source of confusion in the lepton flavor conserving case, due to incomplete information given in most of the papers on the subject. We were able to confirm the results of ref.~\cite{Gratrex:2015hna}, where full and unambiguous information was provided, by an independent explicit calculation. 

\vskip 2.5cm

\section*{Acknowledgments}
We would like to thank Roman Zwicky for his patience in comparing the details of our calculations with their results presented in ref.~\cite{Gratrex:2015hna}, 
and Javier Virto for giving us additional information regarding the results of their global fits presented in ref.~\cite{Descotes-Genon:2015uva}. We thank Michel Davier for discussion related to the leptonic $\tau\to \mu\bar \nu_\mu\nu_\tau$ decay. R.Z.F. thanks the Universit\'e Paris Sud for the kind hospitality, and CNPq and FAPESP for partial financial support. 
We kindly acknowledge support from the European ITN project
(FP7-PEOPLE-2011-ITN, PITNGA-2011-289442-INVISIBLES).

\newpage 
\section*{ Appendix: Angular conventions and Kinematics}

\begin{figure}[h!]
\begin{center}
\includegraphics[scale=0.28]{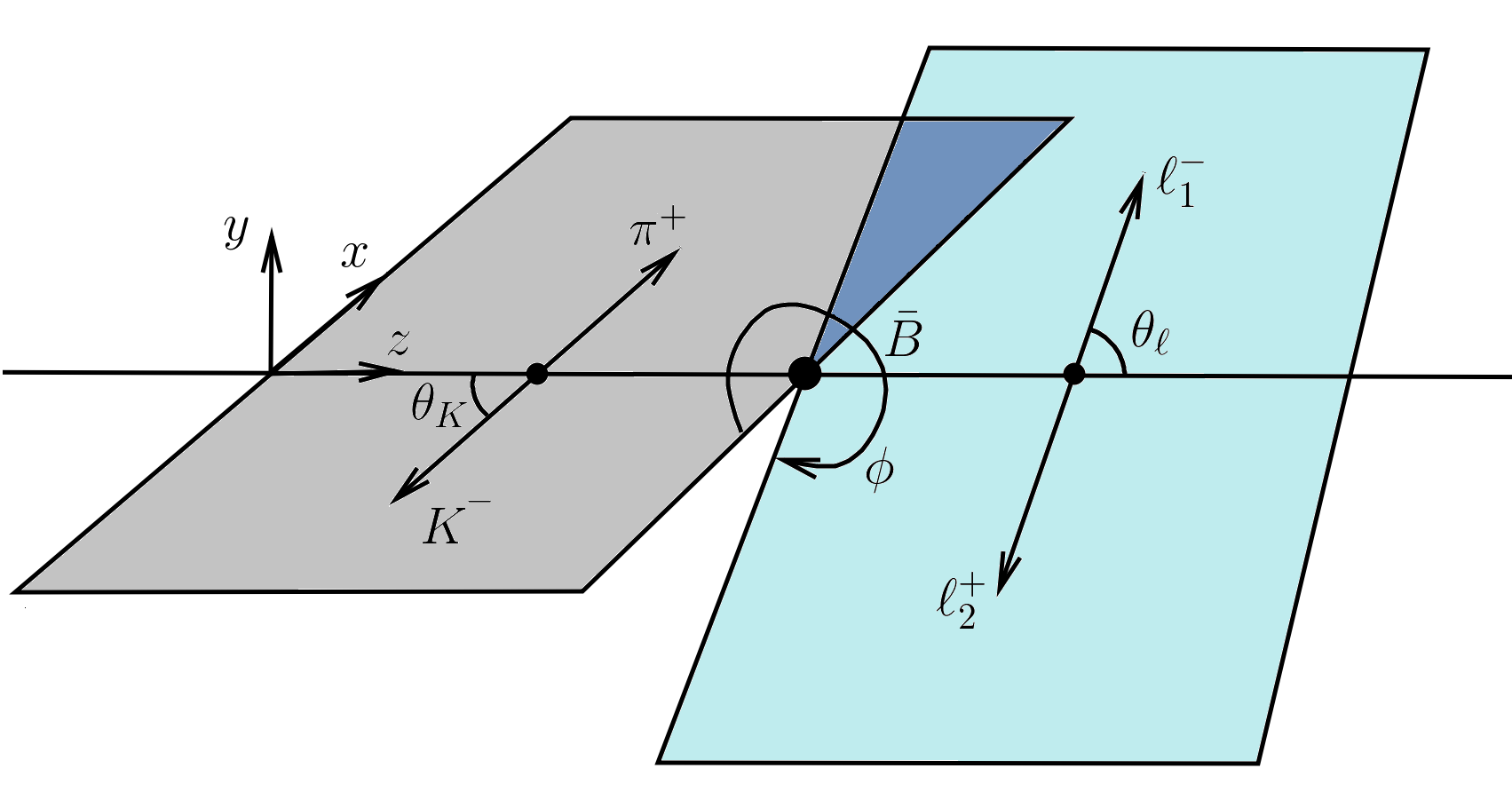}
\end{center}
\caption{\label{fig:angles} \small \sl Angular conventions for the decay $\bar{B}\to \bar{K}^\ast \ell_1^- \ell_2^+$.}
\end{figure}
Our angular conventions for the decay $\bar{B}\to\bar{K}^\ast(\to K^- \pi^+)\ell_1^-\ell_2^+$ are summarized in Fig.~\ref{fig:angles}. 
In the $B$ rest frame, the leptonic and hadronic four-vectors are defined by $q^\mu=(q_0,0,0,q_z)$ and $k^\mu=(k_0,0,0,-q_z)$, where 
\begin{equation}
q_0=\frac{m_B^2+q^2-m_{K^\ast}^2}{2 m_B},\qquad k_0=\frac{m_B^2+m_{K^\ast}^2-q^2}{2 m_B},\quad\text{and}\quad q_z=\frac{\lambda^{1/2}(m_B,m_{K^\ast},\sqrt{q^2})}{2 m_B},
\end{equation}
In the dilepton rest frame, the leptonic four-vectors read
\begin{align}
p_{1}^\mu &= (E_\alpha, |p_\ell| \sin\theta_\ell \cos\phi,-|p_\ell| \sin \theta_\ell \sin\phi,|p_\ell|\cos\theta_\ell),\\
p_{2}^\mu &= (E_\beta, -|p_\ell| \sin\theta_\ell \cos\phi,|p_\ell| \sin \theta_\ell \sin\phi,-|p_\ell|\cos\theta_\ell),
\end{align}
where 
\begin{equation}
E_1=\frac{q^2+m_1^2-m_2^2}{2 \sqrt{q^2}},\qquad E_2=\frac{q^2-m_1^2+m_2^2}{2 \sqrt{q^2}},\quad\text{and}\quad |p_\ell|=\frac{\lambda^{1/2}(q^2,m_1^2,m_2^2)}{2 m_B}.
\end{equation}
In the same way, one can write in the $K^\ast$ rest frame
\begin{align}
p_K^\mu &= (E_K, -|p_K|\sin \theta_K,0,-|p_K|\cos\theta_K),\\
p_\pi^\mu &= (E_\pi,+|p_K|\sin \theta_K,0,+|p_K|\cos\theta_K),
\end{align}
where $E_K$, $E_\pi$ and $|p_K|$ are given by the similar expressions.

\subsection*{Polarization vectors}
In the $B$ rest frame, we choose the polarization vectors to be:
\begin{align}
\varepsilon_V^\mu(\pm) &=\frac{1}{\sqrt{2}}(0,\pm 1,i,0), \hspace{2.1cm} \varepsilon_{K^\ast}^\mu(\pm) = \frac{1}{\sqrt{2}}(0,\mp 1, i,0),\\
\varepsilon_V^\mu(0) &=\frac{1}{\sqrt{q^2}}(q_z,0,0,q_0), \hspace{2.0cm} \varepsilon_{K^\ast}^\mu(0)=\frac{-1}{\sqrt{k^2}}(k_z,0,0,k_0), \nonumber\\
\varepsilon_V^\mu(t) &= \frac{1}{\sqrt{q^2}}(q_0,0,0,q_z), \nonumber
\end{align}
where $V$ stands for for the virtual gauge boson, $Z^\ast$ or $\gamma^\ast$. These four-vectors are orthonormal and satisfy the completeness relations
\begin{align}
\sum_{n,n^\prime}\varepsilon_V^{\ast \mu}(n)\varepsilon_V^\nu(n^\prime) g_{n n^\prime} &= g^{\mu\nu},\\
\sum_{m,m^\prime}\varepsilon_{K^\ast}^{\ast \mu}(m) \varepsilon_{K^\ast}^\nu(m^\prime) \delta_{m m^\prime} &= - g^{\mu\nu}+\frac{k^\mu k^\nu}{m_{K^\ast}^2},
\end{align}
where $m\in\lbrace 0,\pm\rbrace$, $n,n^\prime\in\lbrace 0,t,\pm\rbrace$, and $g_{nn^\prime}=\mathrm{diag}(1,-1,-1,-1)$.

\end{document}